\documentclass{pasj00}

\begin{document}
\SetRunningHead{K. Morokuma-Matsui et al.}{Tight anti-correlation between molecular gas fraction and $4000$~\AA~break strength}

\title{CO emissions from optically selected galaxies at $z\sim0.1-0.2$:\\Tight anti-correlation between molecular gas fraction and $4000$~\AA~break strength}

\author{Kana \textsc{Morokuma-Matsui}} %
\affil{Nobeyama Radio Observatory, 462-2 Nobeyama, Minamimaki-mura, Minamisaku-gun, Nagano 384-1305}
\email{kana.matsui@nao.ac.jp}

\author{Junichi \textsc{Baba}}
\affil{Earth-Life Science Institute, Tokyo Institute of Technology, 2-12-1-IE-29 Ookayama, Meguro-ku, Tokyo, 152-8550}
\email{babajn@elsi.jp}

\author{Kazuo \textsc{Sorai}}
\affil{Department of Physics/Department of Cosmosciences, Hokkaido University, Sapporo, Hokkaido 060-0810}
\email{sorai@astro1.sci.hokudai.ac.jp}

\and
\author{Nario {\sc Kuno}}
\affil{Faculty of Pure and Applied Sciences, University of Tsukuba, 1-1-1, Tennoudai, Tsukuba, Ibaraki 350-8577}
\email{kuno.nario.gt@u.tsukuba.ac.jp}

%

\KeyWords{Galaxies: evolution -- Galaxies: ISM} 

\maketitle

\begin{abstract}

We performed $^{12}$CO($J=1-0$) (hereafter, CO) observations towards 12 normal star-forming galaxies with stellar mass of $M_\star=10^{10.6}-10^{11.3} M_\odot$ at $z=0.1-0.2$ with the 45-m telescope at the Nobeyama Radio Observatory (NRO).
The samples are selected with D$_n(4000)$ that is a strength of the $4000$~\AA~break, instead of commonly used far-infrared (FIR) flux.
We successfully detect the CO emissions from eight galaxies with signal-to-noise ratio (S/N) larger than three, demonstrating the effectiveness of the D$_n(4000)$-based sample selection.
For the first time, we find a tight anti-correlation between D$_n(4000)$ and molecular gas fraction ($f_{\rm mol}$) using literature data of nearby galaxies in which the galaxies with more fuel for star formation have younger stellar populations.
We find that our CO-detected galaxies at $z\sim0.1-0.2$ also follow the same relation of nearby galaxies.
This implies that the galaxies evolve along this D$_n(4000)-f_{\rm mol}$ relation, and that D$_n(4000)$ seems to be used as a proxy for $f_{\rm mol}$ which requires many time-consuming observations.
Based on the comparison with the model calculation with a population synthesis code, we find that star formation from metal enriched gas and its quenching in the early time are necessary to reproduce galaxies with large D$_n(4000)$ and non-zero gas fraction.

\end{abstract}

\section{Introduction}

The redshift range of $z<1$ is the key epoch for revealing formation and evolution of disk galaxies.
Large surveys in the optical and near infrared (NIR) wavelengths have revealed important observational evidence for the dependence of galaxy evolution on its stellar mass:
1) the bimodal distribution in color with the boundary mass of $M_\star=10^{10.5}M_\odot$ (e.g., \cite{Strateva+2001, Kauffmann+2003}),
2) galaxies with $M_\star<10^{10.5}M_\odot$ and $M_\star>10^{10.5}M_\odot$ consist of the later- and earlier-type galaxies, respectively (e.g., \cite{Conselice2006}),
3) massive galaxies with $M_\star>10^{11}M_\odot$ acquired most of their stellar mass before $z=1$ while less massive galaxies with $M_\star<10^{10}M_\odot$ acquired most of their stellar mass after $z=1$ (e.g., \cite{Leitner2012}),
4) for Milky-Way size galaxies, only galactic disks have evolved in $z<1$ and both bulges and disk have been formed by $z>1$ (e.g., \cite{vanDokkum+2013}).
These indicate that the redshift range of $z<1$ is the growing period for disk galaxies in terms of stellar mass.
In addition, local disk galaxies have evolved their galactic disks in this epoch.

Understanding of the stellar component of galaxies at $z<1$ has progressed but there is a very small number of studies on molecular gas of normal galaxies at the redshift range\footnote{
There are studies on Ultra Luminous InfraRed Galaxies (ULIRGs) in this redshift range \citep{Solomon+1997,Chung+2009,Combes+2013}), showing that both the molecular gas fraction and star-formation efficiency play an important role in cosmic SFR evolution \citep{Combes+2013}.
However, most of them have signs of merging (e.g., close pair of galaxies and tidal features), thus they are not expected to be normal galaxies.
}
(\cite{Geach+2011,Matsui+2012,Bauermeister+2013}).
The next most abundant molecule, carbon monoxide (hereafter, CO) after H$_2$ has been widely utilized to measure molecular gas mass since the H$_2$ molecule does not radiate line emission in a cold environment such as molecular clouds whose typical temperature is a few tens Kelvin.
One of the reasons for the small number of CO observations in $z<1$ is the limitation from the atmospheric window. 
The so-called 3-mm window restricted by O$_2$ of atmosphere of the earth is opened around $\sim80-115$ GHz, which constrains the redshift of $z<0.44$ in the case of observations in CO($J=1-0$) whose rest-frame frequency is 115.271 GHz.

The sample selection with far infrared (FIR) flux also prevents the number of CO observations toward normal galaxies from increasing in $z<1$.
For the time-consuming CO observations, it is preferred to observe galaxies with the coordinate, spectroscopic redshift and expected intensity.
Although the catalogue of a large optical survey contains the data of coordinate and redshift of many galaxies (e.g., $\sim10^9$ for SDSS Date Release 9, \cite{Ahn+2012}), the number drastically decreases if the candidates are crossmatched with FIR catalogue.
In addition, sample selection based on FIR flux obviously biases to IR bright objects such as Ultra Luminous InfraRed Galaxies (ULIRGs) whose IR luminosities are $L_{\rm IR}>10^{12}L_\odot$ \citep{SandersMirabel1996}.
The majority of local ULIRGs are reported as mergers with tidal features \citep{Sanders+1988}.
Therefore, to select normal galaxies, sample selection based on FIR flux is not an optimal choice.

We performed $^{12}$CO($J=1-0$) (hereafter, CO) observations toward normal galaxies at $z\sim0.1-0.2$ with the 45-m telescope at the Nobeyama Radio Observatory (NRO)\footnote{Nobeyama Radio Observatory is a branch of the National Astronomical Observatory of Japan, National Institutes of Natural Sciences.} to fill the observational gap in $0.1<z<1.0$.
For the sample selection, we use optical data instead of FIR data to increase the number of candidates and select normal galaxies rather than IR-bright galaxies.

The structure of this paper is as follows:
The observational information and data reductions are described in section \ref{sec:Observations}.
The obtained CO spectra and the relations between molecular gas fraction and optical indices are shown in section \ref{sec:Results}.
We investigate the relation between the star-formation history and gas fraction with model calculation in section \ref{sec:Discussion}.
Finally, we summarize this work in section \ref{sec:Summary}.
We adopt $H_0=71$ km s$^{-1}$, $\Omega_{\rm M}=0.27$ and $\Omega_{\rm M}=0.73$.

\section{Observations}
\label{sec:Observations}

\subsection{Sample Selection}
\label{subsec:SampleSelection}

Sample galaxies are selected using SDSS DR9 \citep{Ahn+2012}, which contains the photometric or spectroscopic data of $\sim10^9$ objects\footnote{http://skyserver.sdss3.org/dr9/en/}.
The basic criteria for the sample selection are as follows:
1) $-20<$ Dec. $<+30$ degree,
2) $0.1<z<0.23$,
3) apparent size: Petrosian radius $>4$ arcsec,
4) nuclear activity: HII,
5) morphology: disk galaxy,
6) D$_n(4000)<1.4$.

The declination range is determined so that the objects are observable with ALMA as well as with the 45-m telescope at NRO for a further investigation of gas distribution and kinematics, which are important for understanding the detailed physical properties of gas component in galaxies.
We put the criterion of apparent disk size to check the morphology of disk galaxy (higher likelihood in fitting of galaxy light profile with exponential than de Vaucouleurs, \cite{{Strateva+2001}}).
Active galactic nuclei (AGN) are excluded from our samples using a set of nebular emission line diagrams (Baldwin, Phillips \& Terlevich, ``BPT'' diagram, \cite{Baldwin+1981}) to reduce the contamination of light from AGN and estimate star-formation rate (SFR) accurately.

\begin{figure}
\includegraphics[width=85mm]{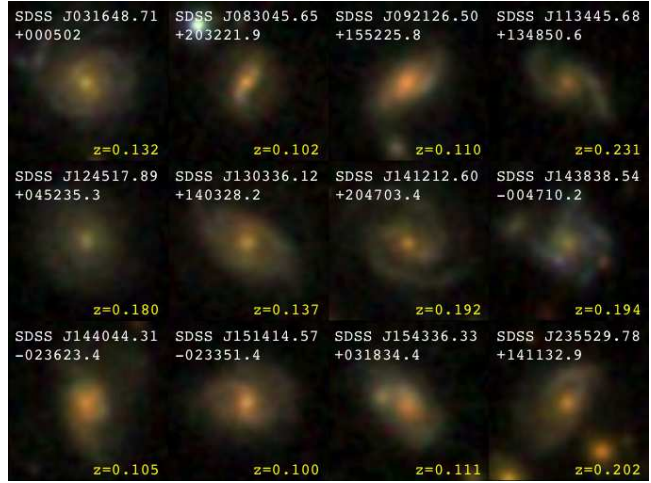}
 \caption{SDSS optical three-color images of the sample galaxies. The image size is $20''\times20''$.}
  \label{fig:SDSSimages}
\end{figure}

We did not use FIR fluxes but D$_n(4000)$ to select sample galaxies.
D$_n(4000)$ is a measure of the strength of 4000~\AA~break and is defined as
\begin{equation}
D_n(4000) = \frac{(\lambda^-_2 - \lambda^-_1) \int^{\lambda^+_2}_{\lambda^+_1} F_\nu\ d\lambda}{(\lambda^+_2 - \lambda^+_1) \int^{\lambda^-_2}_{\lambda^-_1} F_\nu\ d\lambda},
\end{equation}
where $(\lambda^-_1,\lambda^-_2,\lambda^+_1,\lambda^+_2)=(3850,3950,4000,4100)$ in \AA~\citep{Balogh+1999}.
This value allows us to constrain the mean stellar age of a galaxy where lower D$_n(4000)$ indicates younger stellar population and vice versa \citep{Kauffmann+2003}.
If we consider single-burst populations, the 4000~\AA~break, which is characteristic of cooler stars with types later than about G0, becomes dominant at older ages.
This is due to the sudden onset of absorption features bluewards 4000~\AA~produced notably by Ca II H and K, Fe I, Mg I, and CN lines.
\citet{Kauffmann+2012} showed that CO-detected galaxies in the COLD GASS project \citep{Saintonge+2011} have lower value of D$_n(4000)$ ($<1.4$).
We use this empirical criteria for the sample selection instead of commonly used FIR flux.
The SDSS images of the sample galaxies are shown in figure \ref{fig:SDSSimages}.

\subsection{Observations with the 45-m telescope at NRO}
\label{subsec:Observation}

We observed 12 disk galaxies at $z \sim 0.1-0.2$ in the CO ($J=1-0$) line during the observing periods of 2012/2013 and 2013/2014 winter seasons in Japan.
The line frequency was shifted to between 93.638 GHz and 104.749 GHz from the rest frequency of 115.271 GHz according to the redshifts of the sample galaxies.
We used a new two-beam, two-polarization, sideband separating SIS receiver, TZ (Two-beam sideband-separating SIS receiver for Z-machine) \citep{Nakajima+2013} and SAM45 (Spectral Analysis Machine for the 45-m telescope) which is a copy of a part of the FX-type correlator for the Atacama Compact Array (ACA).
The image rejection ratios (IRRs) of TZ were measured at each observing frequency in every observation.
The values of IRRs were $8 - 20$ dB at the center of the intermediate frequency (IF) through the observations.
The system noise temperature, $T_{\rm{sys}}$, was typically $150 - 240$ K during our observations.
Telescope pointing was checked every 50 minutes through observing SiO maser sources close to our target galaxies at 43 GHz.
We used the data that were observed with pointing accuracy better than 5 arcsec.
The main beam size, $\theta_{\rm{B}}$ is $\sim18$ arcsec around 100 GHz, which corresponds to $\sim31$ kpc at the lowest redshift of sample galaxies ($z=0.10$), and to $\sim66$ kpc at the highest ($z=0.23$).
All data were calibrated by the standard chopper wheel method, and converted from antenna temperature $T^*_a$ scale into main-beam brightness temperature by $T_{\rm{mb}}=T^*_a/\eta_{\rm{mb}}$.
The main beam efficiencies, $\eta_{\rm{mb}}$ in the 2012/2013 and 2013/2014 seasons at the observing frequencies were $0.33-0.35$ and $0.37-0.41$, respectively.

\subsection{Data reduction}
\label{subsec:DataReduction}

Data reduction is performed using the NEWSTAR software, which was developed by NRO based on the Astronomical Image Processing System (AIPS) package. 
Integration time of each galaxy is typically six hours summed up for the both polarizations, and the typical r.m.s.~noise temperature was in the range of $1-8$ mK in $T_{\rm{mb}}$ scale after binning up to $40$ $\rm km\ s^{-1}$ resolution.
Integrated intensity, $I_{\rm{CO}}$, is calculated according to $I_{\rm{CO}} = \int T_{\rm{mb}}\ dv$.
The error of $I_{\rm{CO}}$, $\Delta I_{\rm{CO}}$, is estimated as $T_{{\rm r.m.s.}} \sqrt{\Delta V_{\rm e} \Delta v}$, where $T_{{\rm r.m.s.}}$ is the r.m.s.~noise temperatures at velocity resolution of $\Delta v$, which is the velocity resolution of final spectrum.
$\Delta V_{\rm e}$ is the full line width within which the integrated intensity is calculated.
In the case where the peak temperature of the emission line has a signal-to-noise ratio (S/N) less than three, the upper limit of 3 $\Delta I_{\rm{CO}}$ was adopted as $I_{\rm{CO}}$.

CO line luminosity, $L_{\rm{CO}}'$ is calculated from the integrated intensity and beam solid angle $\Omega_{b}=\pi \theta_{\rm{B}}^2/4 \ln{2}$ ($\theta_{\rm{B}}$ in radians) as
\begin{equation}
L_{\rm{CO}}' = \frac{\Omega_{b} I_{\rm{CO}} D_L^2}{(1+z)^3}\ \ (\rm{K\ km\ s^{-1}\ pc^2}),
\end{equation}
where $D_L$ is a luminosity distance. 

Molecular gas mass, $M_{\rm mol}$, is calculated with $L_{\rm CO}'$ and a CO-to-H$_2$ conversion factor, $\alpha_{\rm CO}$ as $M_{\rm mol} = 1.36\ \alpha_{\rm CO} L_{\rm CO}'$,
where 1.36 is a factor to account for the contribution of He by mass.
We adopt the typical Galactic value of $\alpha_{\rm CO}=4.3$ $M_\odot$ (K km s$^{-1}$ pc$^2$)$^{-1}$ (corresponding to $X_{\rm CO}=2\times10^{20} {\rm cm^{-2} [K\ km\ s^{-1}]^{-1}}$), which is widely used for star-forming galaxies in the high-$z$ universe \citep{Tacconi+2013} as well as local normal galaxies \citep{Bolatto+2013}.

\section{Results}
\label{sec:Results}

\subsection{CO spectra}

\begin{figure*}
\begin{center}
\includegraphics[width=140mm]{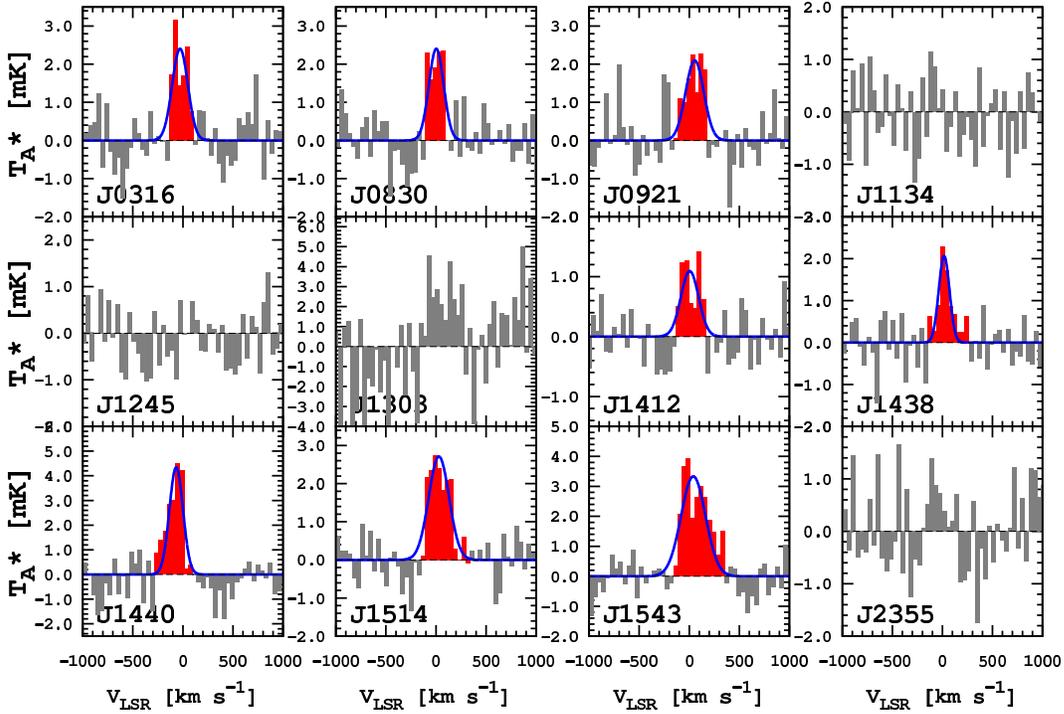}
\end{center}
 \caption{CO spectra obtained with the 45-m telescope at NRO.
 The unit of the vertical axis is antenna temperature $T_a\ast$ and the velocity resolution is $40$ km s$^{-1}$.
 $I_{\rm CO}$ is calculated by summing the intensity within the velocity range colored in red.
 The fitting results with a Gaussian are also shown in blue lines in case of S/N$\ge3$.}
  \label{fig:Spectra}
\end{figure*}

Figure \ref{fig:Spectra} shows the obtained CO spectra binned with a velocity resolution of $40$ km s$^{-1}$.
We successfully detect CO emission from six galaxies with S/N (peak temperature) larger than four, and tentatively from two galaxies ($3\le$ S/N $<4$) while the remaining four galaxies do not show significant CO emission (S/N $<3$).
This high detection rate (67 \%) indicates the effectiveness of the sample selection based on D$_n(4000)$.
The fitting result with a Gaussian is also plotted with blue solid line for the emission lines with S/N $\ge3$.
$I_{\rm CO}$, full width at half maximum (FWHM), molecular gas mass ($M_{\rm mol}$) and molecular gas fraction ($f_{\rm mol}$, equation (\ref{eq:fmol})) are presented in table \ref{tab:Results}.
$I_{\rm CO}$ is calculated by summing the intensity within the velocity range colored in red in figure \ref{fig:Spectra}.
FWHM in this table is estimated with a Gaussian fitting.
Although the line profiles of the galactic CO spectra are not well described by a Gaussian as in figure \ref{fig:Spectra}, we here present the FWHM for reference, which are not used in the following analysis.

\begin{table*}
 \begin{minipage}{\textwidth}
\begin{center}
\caption{General information and observed results of the sample galaxies.}
\begin{tabular}{lcccccccc}
\hline\hline
SDSS Name & $z$ & $M_\star$\footnotemark[$\ast$] & SFR\footnotemark[$\ast$] & D$_n(4000)$\footnotemark[$\ast$] & $I_{\rm CO}$ & FWHM & $M_{\rm mol}$ & $f_{\rm mol}$\\
& & {\footnotesize $\log{(M_\star/M_\odot)}$} & {\footnotesize ($M_\odot/$yr)} & &  {\footnotesize (K km s$^{-1}$)} & {\footnotesize (km s$^{-1}$)} & {\footnotesize ($10^{9} M_\odot$)} & {\footnotesize (\%)}\\
\hline
J031648.71$+$000502.5 & 0.1315 & 10.79 & 10.8 & $1.27\pm0.02$ & $1.35\pm0.18$ & $185\pm53$ & $11.8\pm1.5$ & $16.1\pm1.8$\\
J083045.65$+$203221.9 & 0.1021 & 10.55 & 16.1 & $1.27\pm0.02$ & $1.14\pm0.14$ & $174\pm50$ & $6.03\pm0.77$ & $14.6\pm1.6$\\
J092126.50$+$155225.8 & 0.1097 & 10.94 & 18.7 & $1.24\pm0.01$ & $1.37\pm0.24$  & $214\pm53$ & $8.32\pm1.47$ & $8.7\pm1.4$\\
J113445.68$+$134850.6 & 0.2310 & 10.88 & 21.8 & $1.25\pm0.03$ & $<0.48$ & $-$ & $<12.8$ & $<14.4$\\
J124517.89$+$045235.3 & 0.1800 & 10.75 & 19.0 & $1.23\pm0.03$ & $<0.67$ & $-$ & $<10.9$ & $<16.2$\\
J130336.12$+$140328.2 & 0.1369 & 10.92 & 8.47 & $1.45\pm0.03$\footnotemark[$\ast\ast$] & $<2.46$ & $-$ & $<23.3$ & $<21.7$\\
J141212.60$+$204703.4 & 0.1925 & 11.09 & 18.0 & $1.32\pm0.02$ & $0.73\pm0.12$ & $209\pm89$ & $13.5\pm2.3$ & $9.9\pm1.5$\\
J143838.54$-$004710.2 & 0.1940 & 10.62 & 13.4 & $1.21\pm0.02$ & $0.76\pm0.16$ & $127\pm30$ & $14.4\pm3.0$ & $25.8\pm4.0$\\
J144044.31$-$023623.4 & 0.1050 & 10.89 & 18.9 & $1.29\pm0.03$ & $2.45\pm0.29$ & $159\pm28$ & $9.18\pm1.10$ & $10.7\pm1.1$\\
J151414.57$-$023351.4 & 0.1005 & 10.73 & 11.0 & $1.26\pm0.02$ & $1.75\pm0.16$ & $240\pm46$ & $8.93\pm0.82$ & $14.3\pm1.1$\\
J154336.33$+$031834.4 & 0.1109 & 10.66 & 21.0 & $1.22\pm0.02$ & $3.28\pm0.25$ & $287\pm84$ & $20.4\pm1.6$ & $30.8\pm1.6$\\
J235529.78$+$141132.9 & 0.2015 & 11.31 & 48.4 & $1.34\pm0.03$ & $<0.76$ & $-$ & $<15.4$ & $<7.0$\\
\hline
\end{tabular}
\label{tab:Results}
\end{center}
\footnotetext[$\ast$]{$M_\star$, SFR and D$_n(4000)$ are retrieved from SDSS SkyServer DR10 (http://skyserver.sdss3.org/dr10/en/).
$M_\star$ is estimated from $z$-band magnitude to characterize the galaxy luminosity and $M/L$ ratio from spectral indices of D$_n(4000)$ and H$\delta$ \citep{Kauffmann+2003,Tremonti+2004}.
SFR is estimated from the attenuation-corrected H$\alpha$ luminosity \citep{Brinchmann+2004}.
The infrared luminosity ($L_{\rm IR}$) inferred from SFR ranges $(5.67-32.1)\times10^{10}L_\odot$ using $\log{{\rm SFR}(M_\odot {\rm yr^{-1}})}=\log{L_{\rm IR}({\rm erg\ s^{-1}})}-43.41$ (equation 12 of \cite{KennicuttEvans2012}), confirming that they are not ULIRGs.}
\footnotetext[$\ast\ast$]{D$_n(4000)$ of this galaxies was listed as $1.30\pm0.02$ in DR9 but was changed to $1.45\pm0.03$ in DR10.}
\end{minipage}
\end{table*}

\subsection{Star-formation history}
\label{subsec:d4000}

\begin{figure*}
\begin{center}
\includegraphics[width=140mm]{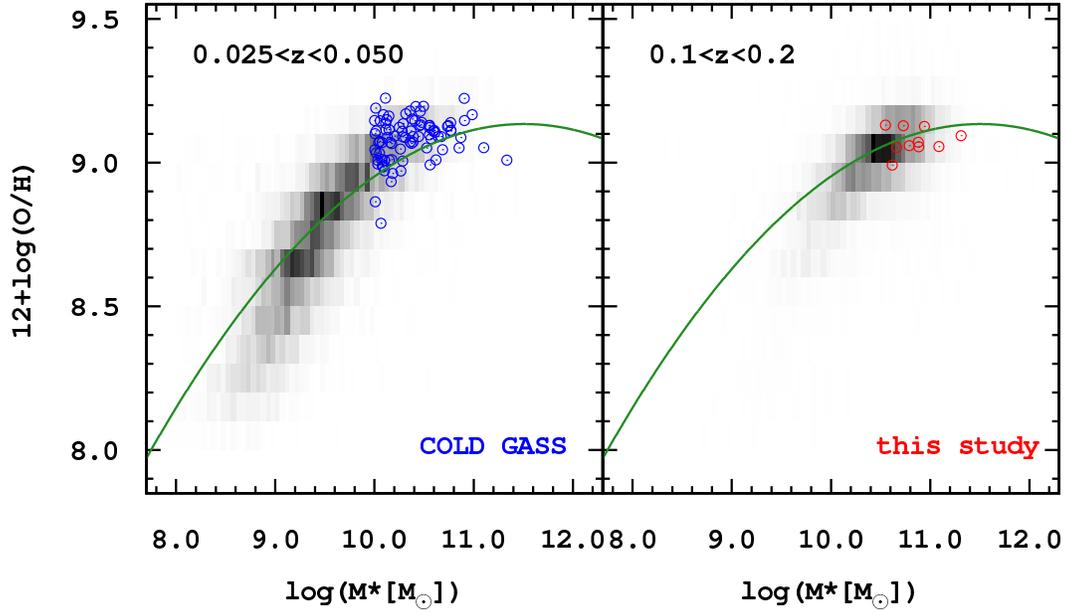}
\end{center}
 \caption{Mass-metallicity relation of COLD GASS samples (\cite{Saintonge+2011}, blue open circles) and our observed galaxies at $z\sim0.1-0.2$ (red open circles).
An empirical mass-metallicity relation of local galaxies obtained in \citet{Tremonti+2004} is shown as green solid line, as $12+\log{({\rm O/H})}=-1.492+1.847(\log{M_\star/M_\odot})-0.08026(\log{M_\star/M_\odot})^2$.
The background grey scale shows the distribution of 42,532 and 51,157 galaxies at $0.025<z<0.050$ and in $0.1<z<0.2$ as a reference.
}
  \label{fig:MZ}
\end{figure*}

\begin{figure*}
\begin{center}
\includegraphics[width=140mm]{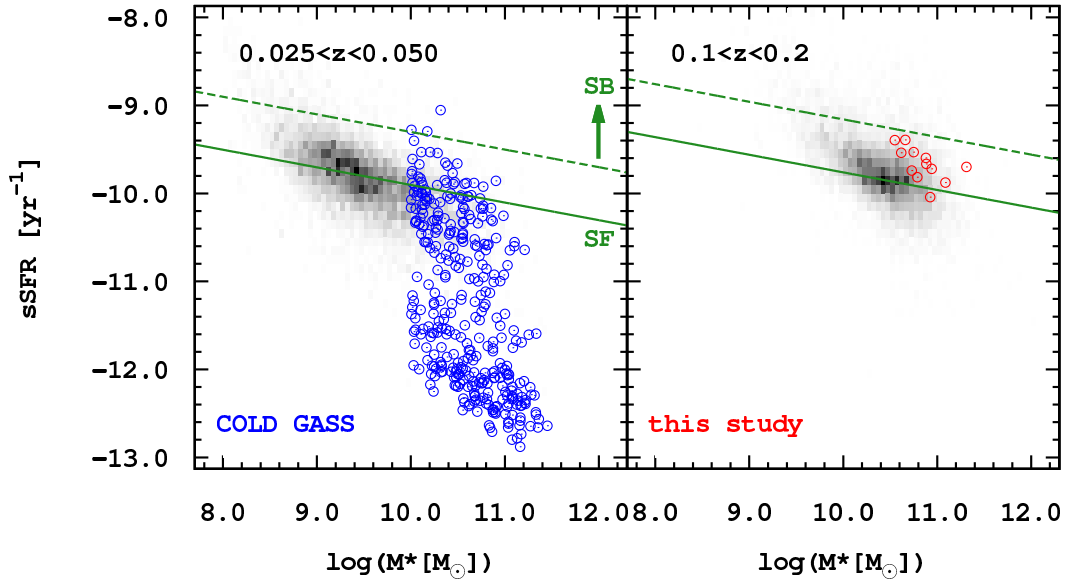}
\end{center}
 \caption{Mass-sSFR relation of COLD GASS samples (\cite{Saintonge+2011}, blue open circles) and our observed galaxies at $z\sim0.1-0.2$ (red open circles).
 An empirical mass-sSFR relation of star-forming galaxies obtained in \citet{Bauermeister+2013} is shown as green solid line, as ${\rm sSFR_{\rm SF} (Gyr^{-1})} = 0.07 (1+z)^{3.2} \left (\frac{M_\star}{10^{11}M_\odot} \right)^{-0.2}$.
 The green dashed line represents boundary between star-forming galaxies and starburst galaxies (${\rm sSFR_{\rm SB}}>4\times{\rm sSFR_{\rm SF}}$).
 The background grey scale shows the distribution of 42,532 and 51,157 galaxies at $0.025<z<0.050$ and in $0.1<z<0.2$ as a reference.
 }
  \label{fig:MsSFR}
\end{figure*}

\begin{figure}
\begin{center}
\includegraphics[width=86mm]{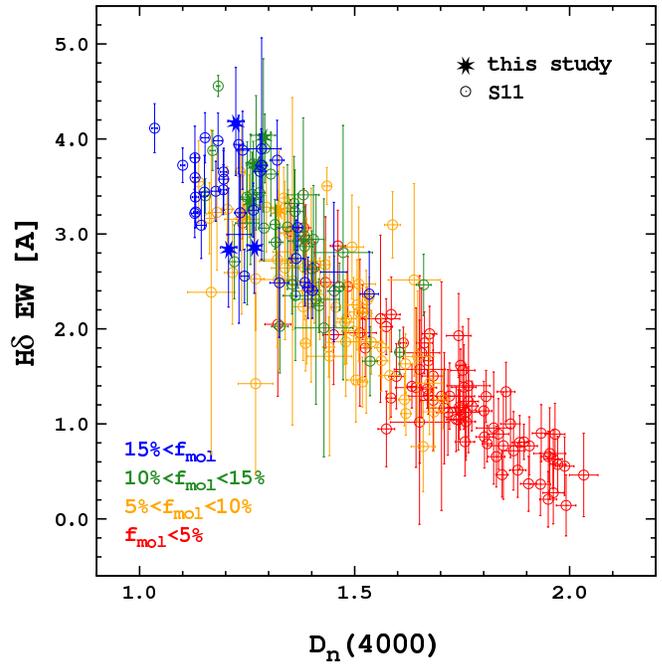}
\end{center}
 \caption{Correlation between D$_n(4000)$ and H$\delta$ EW.
 Observed galaxies at $z\sim0.1-0.2$ and local galaxies (COLD GASS, \cite{Saintonge+2011}) are shown as filled and open circles.
The color is determined according to $f_{\rm mol}$ in which $f_{\rm mol}>15$ \% as blue, $10<f_{\rm mol}<15$ \% as green, $5<f_{\rm mol}<10$ \% as orange and $f_{\rm mol}<5$ \% as red.}
  \label{fig:d4000_hdelta}
\end{figure}

An optical spectrum of a galaxy is the sum of the continuum and absorption spectra of stellar components and nebular emissions from ionized gas.
In addition, stellar spectrum is different according to the stellar type.
Stellar types that dominate the galactic optical light change with time and then the characteristic spectral features accordingly change with time (e.g., \cite{FiocRocca-Volmerange1997,BruzualCharlot2003}).
Therefore, previous studies have investigated the mean stellar ages and star-formation history of galaxies with those characteristic spectral features.

A tight anti-correlation between D$_n(4000)$ and H$\delta$ absorption has been studied to investigate the mean stellar age and star formation history of galaxies (e.g., \cite{Balogh+1999,Kauffmann+2003}).
Measuring the strength of hydrogen Balmer lines in the galactic spectrum is one of the standard methods to derive luminosity-weighted mean ages from the integrated light of galaxies.
Once O- and B-type stars have completed their evolution, Balmer lines become most outstanding in A-type star and weaken as a stellar population gets older.
\citet{Kauffmann+2003} investigated the mean stellar age and star-formation history statistically using SDSS data combined with the model calculations.
They showed that D$_n(4000)$ constrains the mean stellar age of galaxies as described above and that the Balmer absorption-line index, H$\delta_{\rm A}$ can constrain the fractional stellar mass formed in starburst events over the past few Gyr.

We retrieved catalogued data of D$_n(4000)$, H$\delta$ equivalent width (EW), stellar mass ($M_\star$), SFR and metallicity ($12+\log{\rm (O/H)}$) of nearby galaxies and our observed galaxies from the SDSS DR10 \citep{Ahn+2014,Brinchmann+2004,Tremonti+2004}.
H$\delta$ EW is used as a measure of the strength of H$\delta$ absorption, where a larger H$\delta$ EW indicates a stronger absorption.
The data of nearby galaxies is retrieved from COLD GASS \citep{Saintonge+2011}, which is the largest unbiased CO survey toward nearby galaxies ($0.025 < z < 0.05$).
Mass-metallicity and mass-specific SFR (${\rm sSFR =SFR}/M_\star$) relations for these samples are shown in figures \ref{fig:MZ} and \ref{fig:MsSFR}, respectively.
Reference data are also plotted in grey scales, which are selected based on the redshift criteria, $0.025<z<0.050$ for COLD GASS sample and $0.1<z<0.2$ for our observed galaxies, and the criteria that their stellar mass, SFR and metallicity are properly calculated (i.e., not -999).
In total, 42,532 and 51,157 galaxies were left as the reference samples.
The green solid lines in figures \ref{fig:MZ} and \ref{fig:MsSFR} are empirical relations presented in \citet{Tremonti+2004} and \citet{Bauermeister+2013}, respectively.
The green dashed line in figure \ref{fig:MsSFR} represents the boundary between star-forming galaxies and starburst galaxies.
In figure \ref{fig:MZ}, COLD GASS samples and our observed galaxies are found in the plateau at high mass range of the mass-metallicity relation.
Therefore, the variety in metallicity among these galaxies is expected to be small.
In figure \ref{fig:MsSFR}, our observed galaxies are all star-forming galaxies whereas COLD GASS samples include quenched galaxies as well.

In figure \ref{fig:d4000_hdelta}, we compare D$_n(4000)$ and H$\delta$ EW of our sample galaxies at $z\sim0.1-0.2$ (filled star) with those of COLD GASS galaxies (open circle), confirming the anti-correlation reported in previous studies \citep{Kauffmann+2003}.
Moreover, the COLD GASS galaxies are distributed in a sufficiently wide range of D$_n(4000)$ for investigation of various star-formation histories of galaxies.

\subsection{Molecular gas fraction}
\label{subsec:fmol}

In figure \ref{fig:d4000_hdelta}, we change the colors of the symbols according to their molecular gas fractions, $f_{\rm mol}<5$ \% as red, 5 \% $<f_{\rm mol}<10$ \% as orange, 10 \% $<f_{\rm mol}<15$ \% as green and $f_{\rm mol}>15$ \% as blue.
Here, molecular gas fraction is defined as,
\begin{equation}
f_{\rm mol} (\%) = \frac{M_{\rm mol}}{M_{\rm mol}+M_\star}\times100.
\label{eq:fmol}
\end{equation}
It is clearly seen that the COLD GASS galaxies with higher H$\delta$ EW and lower D$_n(4000)$ have higher gas fraction.
In other words, more gas-rich galaxies have younger stellar components compared to the gas-poor galaxies.
Our sample galaxies at $z\sim0.1-0.2$ seem to follow the same D$_n(4000)-$H$\delta$ EW $-f_{\rm mol}$ relation.
This clear trend indicates the existence of the relation between D$_n(4000)$ and $f_{\rm mol}$.

\begin{figure}
\begin{center}
\includegraphics[width=86mm]{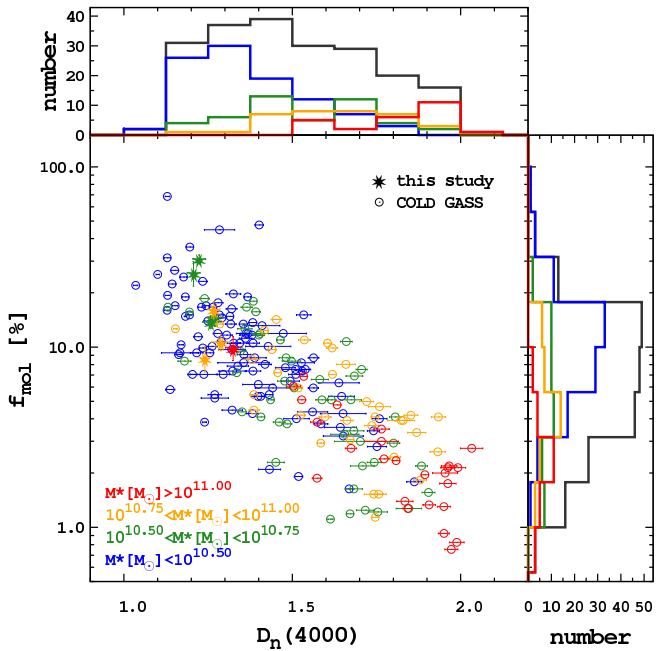}
\end{center}
 \caption{Correlation between D$_n(4000)$ and $f_{\rm mol.}$
 The symbols are the same as figure \ref{fig:d4000_hdelta} but the color is determined according to stellar mass in which $M_\star [M_\odot] >10^{11}$ as red, $10^{10.75}<M_\star [M_\odot]<10^{11}$ as orange, $10^{10.50}<M_\star [M_\odot]<10^{10.75}$ as green and $M_\star [M_\odot]<10^{10.5}$ as blue.}
  \label{fig:d4000_fh2}
\end{figure}

Figure \ref{fig:d4000_fh2} shows the relation between D$_n(4000)$ and $f_{\rm mol}$ of our samples with CO detection (S/N $>3$) and the COLD GASS galaxies.
The symbols are the same as figure \ref{fig:d4000_hdelta}.
As expected, a tight anti-correlation between D$_n(4000)$ and $f_{\rm mol}$ is seen in this figure, suggesting that D$_n(4000)$ could be used as a proxy for $f_{\rm mol}$, which requires much telescope time to be measured.
Since D$_n(4000)$ is calculated as a ratio of fluxes with 100 \AA~widths, D$_n(4000)$ is easily measured compared to the normal spectral line indices such as H$\delta$ EW that require higher spectral resolution and S/N of each spectral channel than D$_n(4000)$.
Therefore, D$_n(4000)$ might be a powerful tool to investigate molecular gas fraction of high-$z$ galaxies
(see also \cite{Tacconi+2013}).

We checked the effect of the difference in the apparent size of galaxies on the D$_n(4000)-f_{\rm mol}$ relation.
D$_n(4000)$ is calculated with data taken within the 3-arcsec fiber in SDSS.
The observed region is different according to the apparent galactic sizes.
Therefore, we compare the apparent size of galaxies and D$_n(4000)$ of the COLD GASS sample.
Even though there is no tight correlation between the apparent size of galaxy and D$_n(4000)$ (correlation factor of 0.07), galaxies larger than 18 arcsec and smaller than 6 arcsec seem to have higher and lower D$_n(4000)$, respectively.
We exclude these larger and smaller samples but confirm the same trend seen in figure \ref{fig:d4000_fh2}.
Hence we conclude that the tight anti-correlation between D$_n(4000)$ and $f_{\rm mol}$ is not affected by the difference in the apparent galactic sizes.

The color of the symbols in figure \ref{fig:d4000_fh2} are determined according to the stellar mass.
The anti-correlation between D$_n(4000)$ and $f_{\rm mol}$ is still seen even if we divide galaxies into four categories according to their stellar masses.
\citet{Kauffmann+2003} reported a bimodal distribution in stellar mass ($M_\star$) $-$ D$_n(4000)$ plot of galaxies, the first peak is found at D$_n(4000)\sim1.3$ and the second peak at D$_n(4000)\sim1.85$.
They also showed that transition mass range ($10^{10}-10^{11}M_\odot$) has a large dispersion in D$_n(4000)$.
Our result suggests that the large dispersion of D$_n(4000)$ seen in $M_\star$ $-$ D$_n(4000)$ is due to the difference in the molecular gas fraction, i.e., the amount of the fuel for the future star formation.
It is notable that this result also indicates the stellar mass-dependent star-formation history is attributed to the difference in the amount of molecular gas in galaxies.
The stellar mass dependence of $f_{\rm mol}$ evolution will be investigated in our forthcoming paper (Morokum-Matsui et al. in prep.).

\begin{figure*}
\begin{center}
\includegraphics[width=140mm]{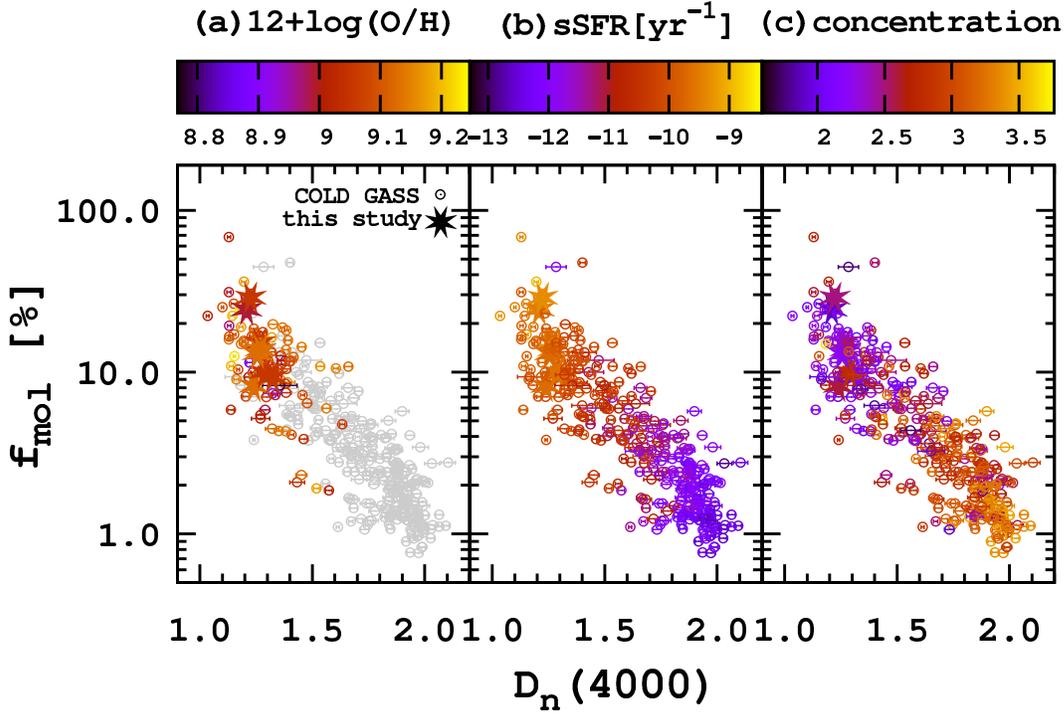}
\end{center}
 \caption{Same as figure \ref{fig:d4000_fh2} but with color coding for (a) gas phase metallicity ($12+\log({\rm O/H})$), (b) specific SFR (${\rm sSFR=SFR}/M_\star$) and (c) concentration parameter ($C=R90/R50$).
Galaxies without metallicity measurement are shown as grey symbols in (a).
 }
  \label{fig:d4000_fmol_ohssfrc}
\end{figure*}

Figure \ref{fig:d4000_fmol_ohssfrc} shows the variations of (a) gas phase metallicity $12+\log{({\rm O/H})}$, (b) sSFR, and (c) concentration parameter in the D$_n(4000)-f_{\rm mol}$ plot.
The concentration parameter, $C$ is defined as the ratio of $C=R90/R50$, where $R90$ and $R50$ are the radii enclosing $90$ \% and $50$ \% of the Petrosian $r$-band luminosity \citep{Petrosian1976} of galaxies \citep{Shimasaku+2001}.
$C$ is strongly related with Hubble type where early- and late-type galaxies have $C\ge2.6$ and $C<2.6$, respectively \citep{Strateva+2001}.
In figure \ref{fig:d4000_fmol_ohssfrc}a, we can see that $12+\log{({\rm O/H})}$ was successfully measured only in star-forming galaxies whose optical spectra contain strong nebular emissions that are used to calculate $12+\log{({\rm O/H})}$.
However, the metallicity variation is expected to be small among COLD GASS and our samples, taking into account that these sample galaxies have stellar mass with $>10^{10}M_\odot$ and that the plateau of mass-metallicity relation is located in $>10^{10}M_\odot$ as well.
We can see the anti-correlation between sSFR and D$_n(4000)$ in figure \ref{fig:d4000_fmol_ohssfrc}b, which was already reported in \citet{Brinchmann+2004}.
Considering that sSFR is a measure of the current versus past star formation, the relation between sSFR and D$_n(4000)$ is not surprising.
$f_{\rm mol}$ is related to sSFR as $f_{\rm mol}=\frac{1}{1+(t_{\rm dep} {\rm sSFR})^{-1}}$ (e.g., \cite{Tacconi+2013}), where $t_{\rm dep}$ is depletion timescale of molecular gas by star formation ($=\frac{M_{\rm mol}}{\rm SFR}$).
If $t_{\rm dep}$ can be assumed to be constant, $f_{\rm mol}$ is a function of sSFR.
Therefore the anti-correlation between D$_n(4000)$ and $f_{\rm mol}$ is also logical conclusion.
Figure \ref{fig:d4000_fmol_ohssfrc}c shows the morphological dependence on D$_n(4000)-f_{\rm mol}$ relation where late-type galaxies tend to have smaller D$_n(4000)$ and larger $f_{\rm mol}$.

\section{Discussion}
\label{sec:Discussion}

\subsection{Comparison with model predictions}

\begin{table*}
\begin{center}
\caption{Common parameters for all template galaxies in PEGASE.2 \citep{FiocRocca-Volmerange1997}.}
\begin{tabular}{lc}
\hline\hline
Parameters & Values \\
\hline
SNII ejecta of massive stars & model B of \citet{WoosleyWeaver1995} \\
Stellar winds & yes \\
Initial mass function & \citet{RanaBasu1992}\\
Lower mass & 0.09 $M_\odot$\\
Upper mass & 120 $M_\odot$\\
Fraction of close binary systems & 0.05\\
Initial metallicity & 0.00\\
Metallicity of the in falling gas & 0.00\\
Consistent evolution of the stellar metallisity & yes\\
Mass fraction of substellar objects & 0.00\\
Nebular emission & yes\\
\hline
\end{tabular}
\label{tab:HubbleSequencePEGASE1}
\end{center}
\end{table*}

\begin{table*}
 \begin{minipage}{\textwidth}
\begin{center}
\caption{PEGASE.2 parameters\footnotemark[$\ast$] for each template galaxy.}
\begin{tabular}{lcccccc}
\hline\hline
Type & p1 & p2 & infall & galactic winds & extinction & age\\
& & {\footnotesize [Myr/$M_\odot$]} & {\footnotesize [Myr]} & {\footnotesize [Gyr]} &  & {\footnotesize [Gyr]}\\
\hline
Burst & -- & -- & -- & -- & inclination-averaged disk geometry & 2\\
E & 1 & 300 & 300 & 1 & spheroidal geometry & 13\\
S0 & 1 & 500 & 100 & 5 & spheroidal geometry & 13\\
Sa & 1 & 1408.5 & 2800 & -- & inclination-averaged disk geometry & 13\\
Sb & 1 & 2500 & 3500 & -- & inclination-averaged disk geometry & 13\\
Sbc & 1 & 5714.3 & 6000 & -- & inclination-averaged disk geometry & 13\\
Sc & 1 & 10000 & 8000 & -- & inclination-averaged disk geometry & 13\\
Sd & 1 & 14286 & 8000 & -- & inclination-averaged disk geometry & 13\\
Im & 1.5 & 15385 & 8000 & -- & inclination-averaged disk geometry & 9\\
\hline
\end{tabular}
\label{tab:HubbleSequencePEGASE2}
\end{center}
\footnotetext[$\ast$]{Burst model assumes $\delta(t)$ for star-formation history and the other models assume SFR as ${\rm SFR}(t)=\frac{1}{\rm p2}M_{\rm gas}(t)^{\rm p1}$.
``infall'' represents the starting time of the gas accretion in Myr.
If there are numbers in the ``galactic winds'' column, the mass of gas in the galaxy becomes zero at the time.
``age'' corresponds to the age of the galaxy at $z=0$.}
\end{minipage}
\end{table*}

\begin{figure*}
\begin{center}
\includegraphics[width=140mm]{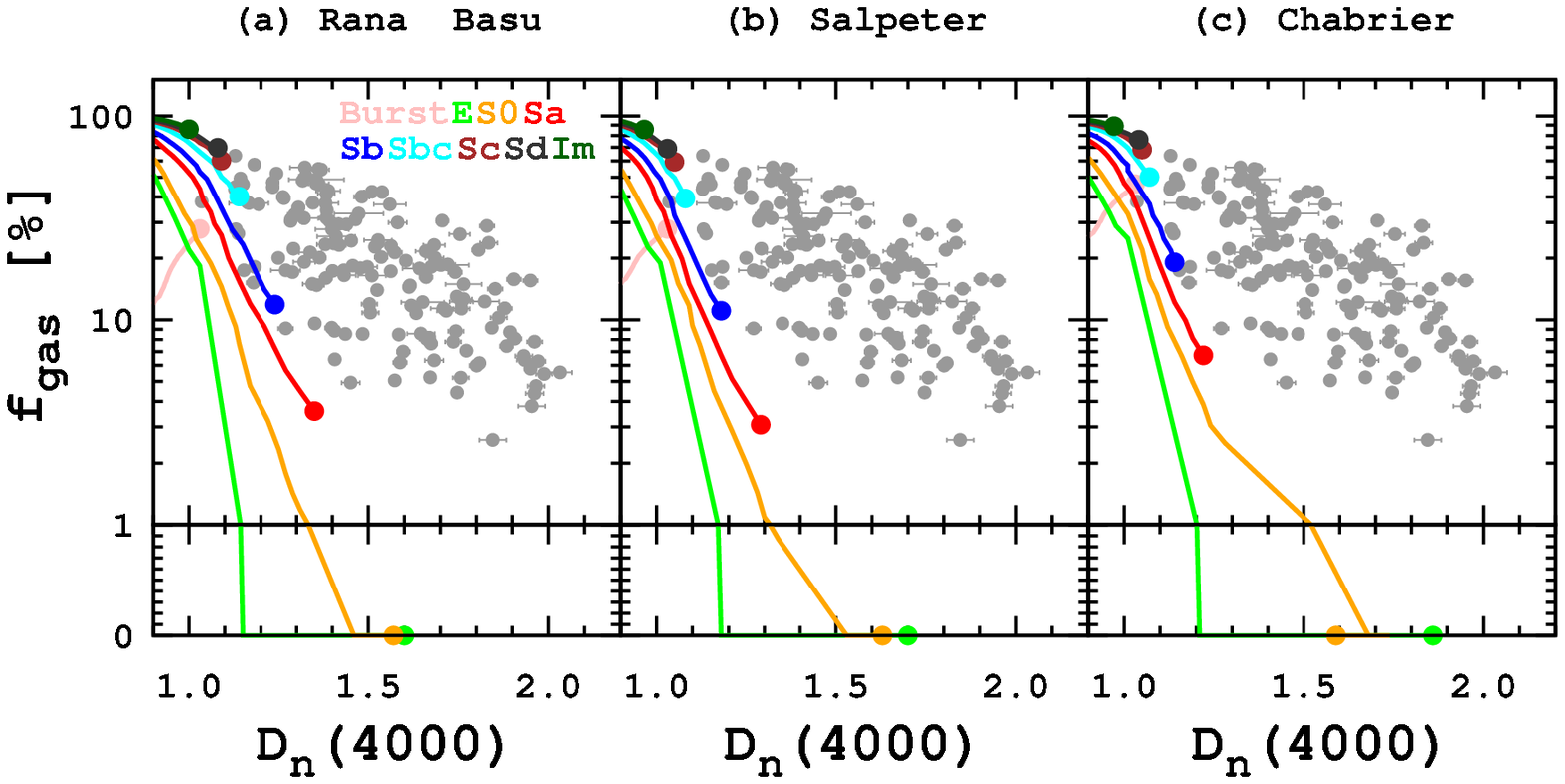}
\end{center}
 \caption{Comparison between the observed D$_n(4000)-f_{\rm mol.}$ relation and the results of the PEGASE calculations in case of the IMF is (a) \citet{RanaBasu1992} (template scenario in PEGASE), (b) \citet{Salpeter1955} and (c) \citet{Chabrier2003}.
 Filled circles represent COLD GASS galaxies from which both CO and HI emissions are detected \citep{Cantinella+2010,Saintonge+2011}. 
 The color-coding of the lines are as follows: Burst in pink, E in light green, S0 in orange, Sa in red, Sb in blue, Sbc in light blue, Sc in brown, Sd in dark brown and Im in green.
 From $f_{\rm gas}=0$ to 1 \%, the plot is in linear scale and from $f_{\rm gas}=1$ to 100 \%, the plot is in log scale.}
  \label{fig:d4000-hdeltafh2_diffIMF}
\end{figure*}

In this section, we compare the observed D$_n(4000)-f_{\rm gas}$ relationship and the evolutionary path in this parameter space predicted by a population synthesis code, PEGASE.2 (\cite{FiocRocca-Volmerange1997}, hereafter PEGASE).
PEGASE allows us to calculate chemical and mass evolution (both gas and stellar components) of galaxies consistently.
We calculated the evolution of nine different types of galaxy (starburst: Burst; early-type galaxies: E, S0; spiral galaxies: Sa, Sb, Sbc, Sc, Sd; irregular galaxies: Im) according to template evolutionary scenarios for them \citep{FiocRocca-Volmerange1997,FiocRocca-Volmerange1999,LeBorgneRocca-Volmerange2002,Tsalmantza+2007}.
In \citet{FiocRocca-Volmerange1999}, the observed spectral energy distributions (SED, from optical to NIR) of $\sim800$ nearby galaxies were used to compute the template SEDs for eight different morphological types (E, S0, Sa, Sb, Sbc, Sc, Sd, Im).
SFRs for the eight types were assumed to be promotional to gas mass as ${\rm SFR}(t)=\frac{1}{\rm p2}M_{\rm gas}(t)^{\rm p1}$.
In addition, SFR of starburst galaxies was modeled to be instantaneous, i.e., $\delta(t)$ \citep{Tsalmantza+2007}.
Exponential gas infall and galactic outflow are also considered.
Galactic outflow is modeled to blow out all the existing gas in a galaxy at the specified time.
\citet{RanaBasu1992} are adopted as the initial mass function (IMF).
In this IMF, the slope in massive star range ($>6M_\odot$) is $\sim-1.7$ that is steeper than Salpeter IMF ($-1.35$, \cite{Salpeter1955}).
The parameters used for the calculation are summarized in tables \ref{tab:HubbleSequencePEGASE1} and \ref{tab:HubbleSequencePEGASE2}.

PEGASE does not calculate evolution of molecular gas but total gas.
Therefore we only use COLD GASS samples combined with available HI data \citep{Cantinella+2010}, since our observed galaxies at $z\sim0.1-0.2$ do not have HI data.
The total gas fraction is calculated as
\begin{equation}
f_{\rm gas}(\%)=\frac{M_{\rm HI}+M_{\rm H_2}}{M_{\rm HI}+M_{\rm H_2}+M_\star}\times 100.
\end{equation}

Figure \ref{fig:d4000-hdeltafh2_diffIMF}a shows the comparison between the observed D$_n(4000)-f_{\rm gas}$ and the results of model calculations for each galaxy type.
Even though the dispersion of the relation is enlarged, there is the same trend in the D$_n(4000)-f_{\rm gas}$ relation of COLD GASS samples, such as one found in figure \ref{fig:d4000_fh2} where galaxies with higher gas fraction have lower D$_n(4000)$.
We can see that the template models reproduce the observed trend where gas-rich galaxies tend to have lower D$_n(4000)$ in figure \ref{fig:d4000-hdeltafh2_diffIMF}a.
In addition, it is also reproduced that the later-type galaxies tend to be distributed in the upper left in this plot (see figure \ref{fig:d4000_fmol_ohssfrc}c).
This qualitative consistency roughly supports the different evolutional scenarios (gas accretion and star-formation timescale) for different types of galaxies proposed in previous study based on optical SED, where later-type galaxies tend to have longer star-formation timescale normalized by total baryonic mass, and longer infall timescale, i.e., more recent gas accretion \citep{FiocRocca-Volmerange1999}.
However there is a significant and qualitative difference between the observations and the model predictions, especially in D$_n(4000)$.
Although the predicted gas fractions in spiral models (Sa $\sim$ Sd) is almost consistent with the observed values, D$_n(4000)$ is too low compared to the observed values.
In early-type galaxies (ETG) models, D$_n(4000)$ evolves to $1.6$ but is still lower than the observed values ($\sim1.9$) for ETG in figure \ref{fig:d4000-hdeltafh2_diffIMF}a.

\subsubsection{Dependences on IMFs}

We also calculated galaxy evolution with the same parameters except for the IMF.
In the previous studies with PEGASE, Rana \& Basu IMF \citep{RanaBasu1992} was used to investigate the other parameters of galaxy evolution, such as star formation law and gas infall timescale, which reproduce observed optical spectra of each type of galaxy \citep{FiocRocca-Volmerange1997,FiocRocca-Volmerange1999,LeBorgneRocca-Volmerange2002,Tsalmantza+2007}.
Figures \ref{fig:d4000-hdeltafh2_diffIMF}b and c show the evolutionary path on the D$_n(4000)-f_{\rm gas}$ plane in cases of two more widely used IMFs, Salpeter IMF \citep{Salpeter1955} and Chabrier IMF \citep{Chabrier2003}, respectively.
Although these two models also fail to reproduce the observed relation, there are slight but important differences among the three models.
First, gas fractions of spiral models at $13$ Gyr with Chabrier IMF are larger than those with Salpeter and Rana \& Basu IMFs.
In addition, D$_n(4000)$ of ETG models at $13$ Gyr with Chabrier IMF are larger than those with Salpeter and Rana \& Basu IMFs. 
These results show that shape of IMF can change the D$_n(4000)-f_{\rm gas}$ relation.
We discuss how the choice of IMF changes the evolution of the D$_n(4000)-f_{\rm gas}$ relation in the following paragraphs.

First of all, the difference in the shape of IMF changes the return gas mass from massive stars when they die.
Compared to less massive stars, massive stars end their lives instantly from their births.
Therefore, the IMF with a larger fraction of massive stars results in a larger gas fraction of galaxy at the time.
The fraction of massive stars is larger in Chabrier IMF than Salpeter and Rana \& Basu IMFs.
This explains the difference seen in the gas fractions of spiral models at $13$ Gyr among three models.

\begin{figure*}
\begin{center}
\includegraphics[width=140mm]{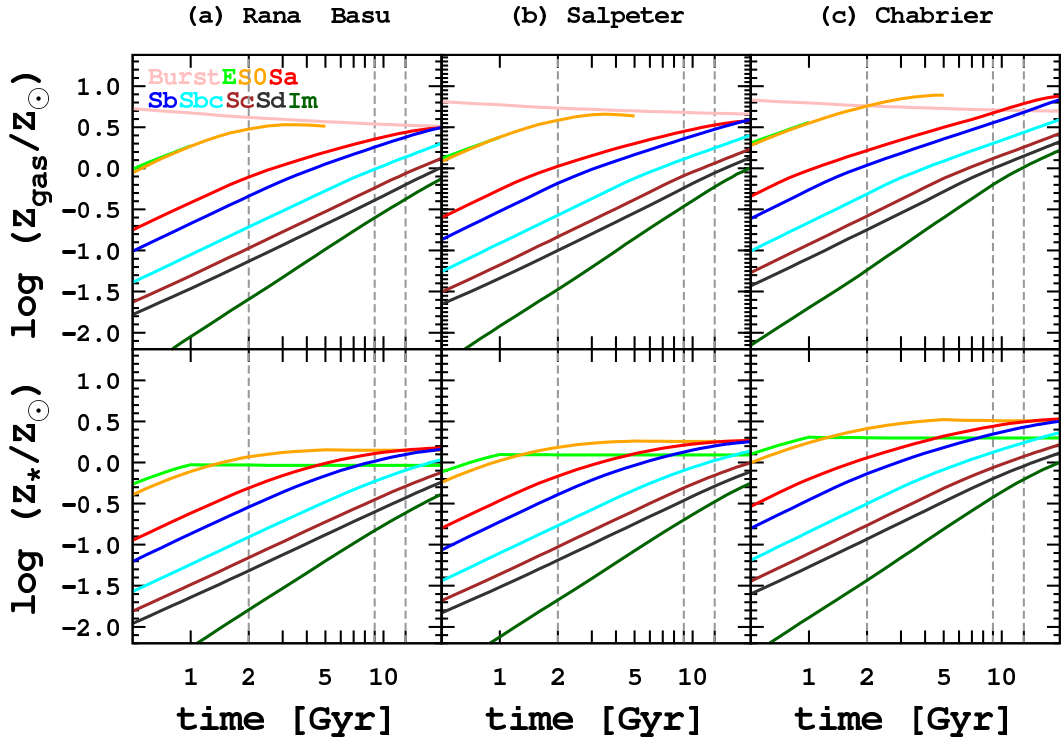}
\end{center}
 \caption{
 Evolution of gas phase (upper panels) and stellar metallicities (lower panels) normalized by the solar value ($Z_\odot=0.0134$, \cite{Asplund+2009}).
 The dashed lines indicate the 2, 9 and 13 Gyrs, which are the ages of each type of galaxy specified in table \ref{tab:HubbleSequencePEGASE2}.
 The color-coding is the same as figure \ref{fig:d4000-hdeltafh2_diffIMF}.
 }
  \label{fig:TemplateZ}
\end{figure*}

Chemical evolution is one of the important factors affecting D$_n(4000)$ evolution.
Current galactic D$_n(4000)$ is determined by the gas phase metallicity when the major portion of stars were formed, and by the elapsed time since the major star-formation event occurred.
It is because 4000 \AA~feature of galactic SED is formed by the metal absorptions of old stellar populations.
\citet{PoggiantiBarbaro1997} investigated the behavior of the 4000 \AA~index for a single star and for a single stellar population (SSP) as a function of age and metallicity.
They explicitly showed that the strength of 4000 \AA~break becomes larger as the metallicity of star and SSP gets large.
Different D$_n(4000)$ evolutions of galaxies according to metallicity were reported in \citet{Kauffmann+2003} where a higher metallicity results in larger D$_n(4000)$ using population synthesis code of \citet{BruzualCharlot2003}.
\citet{Kriek+2011} reported a relation between D$_n(4000)$ and H$\alpha$ EW and showed that the observed relation is reproduced by the code of \citet{BruzualCharlot2003} with a fixed metallicity of solar value through the evolution.
Therefore, it is important to enlarge the metallicity of gas from which the stars are formed to increase D$_n(4000)$ of stars to the observed values.

The difference in the shape of IMF changes the chemical evolution.
The gas released from the dead stars is metal enriched, thus the IMF with a larger fraction of massive stars results in earlier evolution of metallicity.
Figure \ref{fig:TemplateZ} shows the evolutions of gas phase and stellar metallicities with different IMFs.
We can see that the model with Chabrier IMF has metal enriched at an earlier time.
Gas phase and stellar metallicities of ETGs respectively become zero and almost constant (but gradually decreased with time) after the galactic wind blows out all the existing gas\footnote{The gradual decrease of stellar metallicity in ETGs models in figure \ref{fig:TemplateZ} is likely to be due to a small amount of accretion of metal-zero gas.}.
Thus, the ETGs model using the IMF with a larger fraction of massive stars results in larger D$_n(4000)$.

Higher metallicity of gas from which a major portion of stars is formed can increase D$_n(4000)$ as described above.
However, at the same time, $f_{\rm gas}$ must go zero at the early epoch to increase D$_n(4000)$ in the existing condition.
In the next subsection, we discuss the possible reasons for the difficulty in reproducing the observed D$_n(4000)-f_{\rm gas}$ relation quantitatively with the models.

\subsection{Possible reasons for the discrepancy between observation and model calculation}

In this section, we discuss possible reasons for the difficulty in reproducing the observed D$_n(4000)-f_{\rm gas}$ relation in both sides of the model calculation and observation.

\subsubsection{From the model-calculation side}

One of the possibilities is the star-formation model in which stars are formed as long as gas exists regardless of the properties of gas component.
According to the studies on star formation of the local universe, SFR is more closely related with dense molecular gas rather than total gas (e.g., \cite{WongBlitz2002, Bigiel+2008}).
Accreting gas is expected to be diffuse and with high temperature from cosmological simulation of galaxy evolution \citep{Duffy+2012}.
However, there is no model on transition from diffuse and hot gas to dense and cold gas in PEGASE.
Once stars are newly formed, D$_n(4000)$ decreases because the luminosity-weighted SED becomes dominated by young and massive stars taking into account the mass-luminosity relation of main-sequence stars as $L\propto M^{3-4}$ (e.g., \cite{Popper1980,Andersen1991,HenryMcCarthy1993}).
To increase D$_n(4000)$, it is necessary to form stars from metal-enriched gas and quench the star formation in the early time of the galaxy evolution.
Therefore it is difficult to reproduce the galaxies with larger D$_n(4000)$ and non-zero gas fraction with the existing conditions in the model.

Morphological quenching (MQ) is an important process that can quench star formation even with a gas component, but which is not implemented in PEGASE.
\citet{Martig+2009} showed that once the central bulge becomes massive enough to stabilize the galactic disk against local gravitational instability, star-formation quenching occurs.
Central concentration of ETGs leads to large epicyclic frequency $\kappa$, which is linked to the depth of the gravitational potential, and results in large Toomre's $Q$ \citep{Toomre1964}.
As a result, the star formation in ETGs is quenched without a removal of gas component.
We showed the concentration parameters of the COLD GASS sample in figure \ref{fig:d4000_fmol_ohssfrc}c.
We can see that galaxies distributed in the lower-right corner in figure \ref{fig:d4000_fmol_ohssfrc} mainly consist of ETGs.
Therefore, our result suggests that they have large D$_n(4000)$ and non-zero gas fraction because MQ may work in those systems.

Re-accretion of the enriched gas in halo is also an important process for galaxy evolution especially in terms of chemical evolution, which is not modeled in PEGASE.
\citet{Shen+2012} investigated the metal-enriched circumgalactic medium (CGM) of a massive galaxy at $z=3$ ($M_\star=2.1\times10^{10}M_\odot$).
They listed three sources of heavy elements in CGM in order of the fraction, 1) galactic wind from the main host, 2) ``satellite progenitors'' accreted by the main host before $z=3$, 3) ``nearby dwarfs'' orbiting outside the virial radius.
According to their result, the metallicity of accreting gas increases with time as a consequence of re-accretion of the metal-enriched gas in a ``halo fountain'' \citep{Oppenheimer+2010}.
Re-accretion of the enriched gas is likely to promote chemical evolution and increase D$_n(4000)$ of modeled spiral galaxies.
In addition, since re-accretion is reported to be more promoted in more massive systems due to the deceleration by the interaction with dense halo gas \citep{OppenheimerDave2008,Oppenheimer+2010}, the larger discrepancy in D$_n(4000)$ of more massive galaxies may be compensated.

\subsubsection{From the observation side}

The results of PEGASE calculation is a galactic D$_n(4000)-f_{\rm gas}$ relation since PEGASE assumes a galaxy as an one-zone system. 
However, the gas mass of galaxies used in this study is a total mass and D$_n(4000)$ is calculated using data obtained within the SDSS 3-arcsec fiber.
Therefore it is not clear whether the obtained D$_n(4000)-f_{\rm gas}$ relation is a relation between galactic star formation history and galactic gas fraction or star-formation history of central bulge and galactic gas fraction.
Therefore, the much larger D$_n(4000)$ obtained with the observation than those in model prediction may be partly attributed to this situation, since most galaxies have metallicity gradients (e.g., \cite{Sanchez+2014}).
Considering that most disk galaxies have exponential profile of molecular gas distribution and most ETGs have molecular gas distribution with a relatively small extent (e.g., \cite{Davis+2013,Alatalo+2013}), the contribution of CO emission from central part to the total CO flux is large in both cases.
Thus it may be a relation indicating the evolution of central bulge.
To understand the physical background of this relation, it is important to compare spatially disaggregated D$_n(4000)$ and gas fraction.

The universality of this relation should be investigated with higher-$z$ galaxies.
Recent CO observations towards galaxies at $z=1-2$ revealed extremely high molecular gas fractions ($\sim50$ \%) in those systems (e.g., \cite{Tacconi+2013}).
The measurement of D$_n(4000)$ of those systems allows us to enlarge the $f_{\rm mol}$ as well as the redshift ranges at the same time.
Although the reported D$_n(4000)-f_{\rm mol}$ relation in this study has already given the important suggestion, once the observational assessments for the tasks described above are completed, this relation would be one of the strict observational constraints on the models for galaxy formation and evolution.

\section{Summary}
\label{sec:Summary}

We observed 12 normal galaxies at $z\sim0.1-0.2$ with the 45-m telescope at NRO to measure the molecular gas mass and investigate the relation between star formation history and molecular gas fraction $f_{\rm mol}$.
The sample galaxies are selected with D$_n(4000)$ instead of the widely used FIR flux.
The main results obtained in this paper are as follows:
\begin{enumerate}
\item We detect the CO emissions from six galaxies with S/N$>4$, two galaxies with $3<$S/N$<4$ out of 12 samples and show the validity of the sample selection based on D$_n(4000)$.
\item Using the literature data of nearby galaxies \citep{Saintonge+2011}, we find a tight anti-correlation between D$_n(4000)$ and $f_{\rm mol}$ indicating that more gas-rich galaxies tend to have younger stellar populations than gas-poor galaxies for the first time.
\item We show that our sample galaxies at $z\sim0.1-0.2$ follow the same D$_n(4000)$ relation of the nearby galaxies.
This suggests that galaxies might evolve along this relation and that D$_n(4000)$ might be used as proxy for $f_{\rm mol}$ which requires much telescope time to be measured.
\item We calculate the galaxy evolution with a population synthesis code PEGASE \citep{FiocRocca-Volmerange1997} to investigate the observed D$_n(4000)-f_{\rm gas}$ relation.
We calculate the total gas fraction $f_{\rm gas}$ with literature data of nearby galaxies and compare with the model calculations.
As a result, any template scenarios for different morphological types, which have been constructed so as to reproduce optical SED, cannot reproduce the observed relation.
\item Our results suggest that star formation from metal enriched gas and star formation quenching in the early time are necessary to form galaxies with the observed large D$_n(4000)$ and non-zero gas fraction.

\end{enumerate}


\bigskip

We would like to thank an anonymous referee for very productive comments.
KMM thanks Kouji Ohta, Tadayuki Kodama, Takashi Okamoto, Yoichi Tamura, Shinya Komugi, Nick Scoville, Tomoki Morokuma and all members of NRO for their support and fruitful discussions.

PEGASE calculations were carried out on computers at the Center for Computational Astrophysics,
National Astronomical Observatory of Japan.
JB was supported by HPCI Strategic Program Field 5 ``The origin of matter and the universe'' and JSPS Grant-in-Aid for Young Scientists (B) Grand Number 26800099.

Funding for SDSS-III has been provided by the Alfred P. Sloan Foundation, the Participating Institutions, the National Science Foundation, and the U.S. Department of Energy Office of Science.
The SDSS-III web site is http://www.sdss3.org/.
SDSS-III is managed by the Astrophysical Research Consortium for the Participating Institutions of the SDSS-III Collaboration including the University of Arizona, the Brazilian Participation Group, Brookhaven National Laboratory, Carnegie Mellon University, University of Florida, the French Participation Group, the German Participation Group, Harvard University, the Instituto de Astrofisica de Canarias, the Michigan State/Notre Dame/JINA Participation Group, Johns Hopkins University, Lawrence Berkeley National Laboratory, Max Planck Institute for Astrophysics, Max Planck Institute for Extraterrestrial Physics, New Mexico State University, New York University, Ohio State University, Pennsylvania State University, University of Portsmouth, Princeton University, the Spanish Participation Group, University of Tokyo, University of Utah, Vanderbilt University, University of Virginia, University of Washington, and Yale University.





\end{document}